\documentclass[12pt]{article}

\usepackage{amsmath,amssymb,amsfonts,amsthm}
\usepackage{algorithmic}
\usepackage{graphicx}
\usepackage{textcomp}
\usepackage{xcolor}
\usepackage{bbm}
\usepackage{optidef}
\usepackage{enumerate}
\usepackage{cleveref}
\crefrangeformat{equation}{(#3#1#4)--(#5#2#6)}
\usepackage{subcaption}
\usepackage{tikz}
\usepackage{pgfplots, pgfplotstable}
\usepgfplotslibrary{groupplots}
\pgfplotsset{compat=newest}
\usetikzlibrary{shapes,arrows,positioning}

\usepackage{authblk}

\usepackage{chngcntr}

\begin{document}
\newcommand{\ones}{\mathbf{1}}
\newcommand{\infnorm}[1]{\lVert{#1}\rVert_{\infty}}
\newcommand{\infinfnorm}[1]{\lVert{#1}\rVert_{\infty,\infty}}
\newcommand{\twotwonorm}[1]{\lVert{#1}\rVert_{2}}
\newcommand{\onenorm}[1]{\lVert{#1}\rVert_{1}}
\newcommand{\twonorm}[1]{\lVert{#1}\rVert_{2}}
\newcommand{\inv}{^{-1}}
\newcommand{\R}{\mathbb{R}}
\newcommand{\hatf}{\hat{f}}
\newcommand{\hath}{\hat{h}}
\newcommand{\sat}[1]{\text{sat}\left({#1}\right)}
\newcommand{\dz}[1]{\text{dz}\left({#1}\right)}
\newcommand{\sign}[1]{\text{sign}\left({#1}\right)}
\newcommand{\F}{$\mathcal{F}$}

\newtheorem{problem}{Problem}
\newtheorem{theorem}{Theorem}
\newtheorem{lemma}{Lemma}
\newtheorem{remark}{Remark}
\newtheorem{definition}{Definition}
\newtheorem{proposition}{Proposition}
\newtheorem{assumption}{Assumption}
\newtheorem{corollary}{Corollary}

\newcommand*{\review}{\textcolor{black}}

\title{Decentralized PI-control and Anti-windup in Resource Sharing Networks}

\author[1]{Felix Agner\thanks{Corresponding author: felix.agner@control.lth.se.}}
\author[1]{Jonas Hansson}
\author[2]{Pauline Kergus}
\author[1]{Anders Rantzer}
\author[3]{Sophie Tarbouriech}
\author[3,4]{Luca Zaccarian}

\affil[1]{Department of Automatic Control, Lund University}
\affil[2]{LAPLACE, Université de Toulouse, CNRS}
\affil[3]{LAAS-CNRS, Université de Toulouse, CNRS}
\affil[4]{Department of Industrial Engineering, University of Trento}

\date{} 
\maketitle

\begin{abstract}
We consider control of multiple stable first-order \review{agents} which have a control coupling described by an M-matrix. These agents are subject to incremental sector-bounded \review{input} nonlinearities. We show that such plants can be globally asymptotically stabilized to a unique equilibrium using fully decentralized proportional-integral controllers equipped with anti-windup and subject to local tuning rules. In addition, we show that when the nonlinearities correspond to the saturation function, the closed loop asymptotically minimizes a weighted 1-norm of the agents state mismatch. The control strategy is finally compared to other state-of-the-art controllers on a numerical district heating example.
\end{abstract}




\newpage
\section{Introduction}
In this paper we consider the control of agents sharing a central distribution system with limited capacity. These are systems where the positive action of one agent negatively impacts others. This type of competitive structure can arise in many domains, for instance internet congestion control \cite{internet_congestion_lowetal, fairness_internet_congestion_kelly} and district heating systems \cite{AGNER2022100067}. In the district heating scenario, the structure arises because of the hydraulic constraints of the grid. If one agent (building) locally decides to increase their heat demand by opening their control valves, this will lead to higher flow rates and greater frictional pressure losses. These losses make it so that other agents now receive lower flow rates \cite{AGNER2022100067}. We consider a simple description of such systems: \review{
\begin{equation}
    \Dot{x} = -Ax + Bf(u) + w.
    \label{eq:system description}
\end{equation}
Here each agent $i$ is associated with a state $x_i$, and these states are gathered in the vector $x$. The agents are subject to an external disturbance $w$ and interconnected via the matrix $B$. The nonlinear function $f(\cdot)$ can for instance represent the common phenomenon of input saturation, which motivates this work. $A$ is assumed diagonal. We will more formally describe the plant in Section \ref{sec: preliminaries}.}

In multi-agent systems such as \eqref{eq:system description}, decentralized controllers are desirable. \review{Semi-decentralized control strategies for multi-agent systems subject to input saturation have been considered in the following works. In \cite{sync_of_identical_saturating_multiagent_systems}, each networked agent is equipped with a local controller that receives the control input of its neighbors. In \cite{decentralized_aw_UAV}, semi-decentralized anti-windup was considered for stable SISO plants that are decentralized in the linear domain, but become coupled during saturation. This is demonstrated on unmanned aerial vehicles. These works focus on stabilization when the disturbance $w$ in plant \eqref{eq:system description} is energy bounded.} In this work we focus instead on the asymptotic properties of plant \eqref{eq:system description}, which become important when $w$ is expected to vary slower than the dynamics of the plant and can be approximated as constant. Previous works considering asymptotic optimality for plants of the form \eqref{eq:system description} are \cite{BAUSO_asymptotical_optimality} and \cite{agner_coortinating_aw}. \review{In \cite{agner_coortinating_aw}, it was shown that, when $B$ is an M-matrix, decentralized PI-controllers with a rank-one coordinating anti-windup scheme can minimize the cost $\max_i |x_i|$}. In \cite{BAUSO_asymptotical_optimality}, it was shown that the static controller $u = -B^\top x$ asymptotically minimizes the cost $x^\top Ax + v^\top v$ where $v = \sat{u}$. This result also extends to the case when $B$ is not an M-matrix. Both of these control strategies maintain certain scalability properties: With $u = -B^\top x$ \cite{BAUSO_asymptotical_optimality}, any sparsity structure in the $B$-matrix is maintained and the rank-one coordination scheme of \cite{agner_coortinating_aw} admits scalable implementations. However, the most scalable control solution is one that is fully decentralized. In this work, we analyze \eqref{eq:system description} under a fully decentralized PI (proportional-integral) control strategy. In general, it is non-trivial that decentralized PI-controllers are stabilizing, let alone fulfill any optimality criterion. \review{However, in this work we show not only that our strategy minimizes asymptotic costs of the form $\sum_{i=1}^n \gamma_i|x_i|$ but also that the resulting equilibrium is globally asymptotically stable under decentralized controller tuning rules.}

The paper is organized as follows: Section \ref{sec: preliminaries} presents the considered plant and control strategy. Section \ref{sec: main results} presents the main results of the paper, namely equilibrium existence and uniqueness, global asymptotic stability, and equilibrium optimality for our considered closed loop. A motivating numerical example consisting in the flow control of a simplified district-heating network is subsequently given in section \ref{sec: example}. The proofs of the main results are presented in sections \ref{sec: equilibrium existence}, \ref{sec: stability}, and \ref{sec:optimality proof} respectively. Conclusions and future work are covered in section \ref{sec: conclusions}.

\textit{Notation:} $v_i$ denotes element $i$ of vector $v \in \R^n$, $A_i$ denotes row $i$ of matrix $A \in \R^{n \times m}$, and $A_{i, j}$ denotes its $(i,j)$-th element. A matrix $A$ is strictly diagonally row-dominant if $|A_{i,i}| > \sum_{j \neq i} |A_{i,j}|$ for all $i$. $A$ is strictly diagonally column-dominant if $A^\top$, denoting the transpose of $A$, is strictly diagonally row-dominant. \review{Matrix $B \in \R^{n \times n}$ is called positive stable if all of its eigenvalues have positive real part. We denote $S \succ 0$ ($S \succeq 0)$ if $S \in R^{n \times n}$ is symmetric and positive definite (semi-definite). Similarly, for two symmetric matrices $S_1$, $S_2 \in R^{n \times n}$ we denote $S_1 \succ S_2$ ($S_1 \succeq S_2$) if $S_1 - S_2 \succ 0$ ($S_1 - S_2 \succeq 0$).} Let the 2-norm of a vector $x \in \R^n$ be given by $\twonorm{x} = (\sum_{i=1}^n x_i^2)^{1/2}$. Let the 1-and-infinity-norms of a vector $x \in \R^n$ be given by $\onenorm{x} = \sum_{i=1}^n |x_i|$ and $\infnorm{x} = \max_i |x_i|$ respectively. Let the norm $\twotwonorm{A}$ of a matrix $A$ be the induced 2-norm. Let $\ones \in \R^n$ be a vector of all ones, where $n$ is taken in context.
We say that a function $f : \R \to \R$ is increasing (non-decreasing) if $y > x$ implies that $f(y) > f(x)$ ($f(y) \geq f(x)$).

\section{Problem Data and Proposed Controller} \label{sec: preliminaries}
We consider control of plants of the form \eqref{eq:system description} where vector $x\in \mathbb{R}^{n}$ gathers the states $x_i$ of each agent, $A \in \mathbb{R}^{n \times n}$, and $w\in \mathbb{R}^{n}$ is a constant disturbance acting on the plant. $B \in \mathbb{R}^{n \times n}$ couples the control-inputs of the agents. The input nonlinearity $f:\R^n \to \R^n$ is characterized by the following assumption.
\begin{assumption}
    $f(x) = \left[ f_1(x_1), f_2(x_2), \dots, f_n(x_n)\right]^\top$ has components $f_i$ satisfying $f_i(0)=0$ and incrementally sector-bounded in the sector $\left[0, 1\right]$, namely satisfying $0 \leq \left( f_i(y)-f_i(x) \right)/\left(y-x\right) \leq 1$ for all $x \in \R$, $y \in \R$, $x \neq y$.
    \label{ass: function assumption}
\end{assumption}
Note that Assumption \ref{ass: function assumption} implies that $f$ is non-decreasing and Lipschitz with Lipschitz constant 1. Since $f(0)=0$, $f$ also enjoys a sector $\left[0, 1\right]$ condition.

Stability properties for feedback with incrementally sector-bounded nonlinearities has long been considered in the literature. As far back as \cite{zames_incremental_conic} it was used for input-output stability analysis. Both \cite{robust_sync_incremental_diagonal_sectors} and \cite{delellis_adaptive_pinning_diagonal_incremental_sector} consider the type of diagonally partitioned incrementally sector-bounded functions that we consider here, whereas \cite{LMI_conditions_contraction_sector_symmetric, distributed_sync_incremental_nonlinearities_symmetric_sector, incremental_sector_LMI_based_Inf_gain_margin_symmetric_sector} consider a richer class of incremental sector-bound constraints $\left(f(x)-f(y)-S_1(x-y)\right)^\top \left( f(x)-f(y)-S_2(x-y)\right) \leq 0$ for all $x \in \R^n$, $y \in \R^n$. Here $S_1$ and $S_2$ are real symmetric matrices with $0 \preceq S_1 \prec S_2$.

We will consider function pairs $f(\cdot)$, $h(\cdot)$ where $f(x) + h(x) = x$. These pairs fulfill the following property, the proof of which is in the appendix.
\begin{lemma}\label{lem: u - f(u) has same properies}
    Let $f: \R^n \to \R^n$ satisfy Assumption \ref{ass: function assumption}. Then $h(u) = u - f(u)$ also satisfies Assumption \ref{ass: function assumption}.
\end{lemma}
The considered class of function pairs is well motivated by the common case $f(x) = \sat{x}$ where $\sat{x} = \max \left( \min \left( x, \ones \right), -\ones \right)$ and $h(x) = \dz{x} = x - \sat{x}$. 

We propose controlling the plant \eqref{eq:system description} with fully decentralized PI controllers having decentralized anti-windup for each agent $i = 1,\dots,n$.
\begin{align}
    \dot{z_i} = x_i + s_i h_i(u_i) \label{eq:controller z}\\
    u_i = - p_i x_i - r_i z_i \label{eq:controller u}
\end{align}
where $z_i$ is the integral state, $u_i$ is the controller output, $p_i > 0$ and $r_i > 0$ are proportional and integral controller gains respectively, $s_i > 0$ is an anti-windup gain, and $h(u) = u - f(u)$ is an anti-windup signal. Note that while the notation $h$ is not needed (indeed we could equivalently replace $h(u)$ with $u-f(u)$), we will use the pair $f$, $h$ both to simplify the exposition and to highlight that $f$ is the nonlinearity acting on the plant while $h$ is the nonlinearity acting on the controller. We assume that the closed loop system satisfies the following assumption.
\begin{assumption} \label{ass: system properties}
    $A$ is a diagonal positive definite matrix, $B$ is an M-matrix, and $w$ is a constant disturbance. The controller parameters $p_i$, $r_i$, and $s_i$, for $i=1,\dots,n$, are all positive.
\end{assumption}
\review{
The M-matrix property which we consider for $B$ has the following standard definition \cite[p. 113]{horn_johnson_1991}.
\begin{definition}
    A matrix $B \in R^{n \times n}$ is called an M-matrix if $B$ is positive stable and all off-diagonal elements of $B$ are non-positive.
\end{definition}
M-matrices hold certain exploitable properties as listed in Theorem 2.5.3 of  \cite[pp. 114-115]{horn_johnson_1991}. We summarize the ones we employ in this paper in the following proposition.
\begin{proposition} \label{prop:M matrices}
If $B \in R^{n \times n}$ has only non-positive off-diagonal elements, then the following statements are equivalent:
    \begin{enumerate}[(i)]
    \item $B$ is positive stable, that is, $B$ is an M-matrix.
    \item $DB$ is an M-matrix for every positive definite diagonal matrix $D$.    
    \item There exists a diagonal positive definite matrix $U$ such that $U B$ and $U B U\inv$ are strictly column-diagonally dominant.
    \item There exists a diagonal positive definite matrix $Q$ such that \mbox{$QB + B^\top Q \succ 0$}.
\end{enumerate}
\end{proposition}
}
\section{Main Results}\label{sec: main results}
In this section we will cover the main results of this paper. In particular, we will consider the proposed control law \crefrange{eq:controller z}{eq:controller u} for the plant \eqref{eq:system description}. We will show that this closed loop system admits an equilibrium for any constant disturbance $w$. We will additionally show that this equilibrium is globally asymptotically stable and enjoys a notion of optimality. We will leave the proofs for Sections \ref{sec: equilibrium existence} to \ref{sec:optimality proof}.

Let us first consider the existence of an equilibrium, which corresponds to well-posedness of the equations \crefrange{eq:system description}{eq:controller u} with $\dot{x} = \dot{z} = 0$.
\begin{theorem}[Equilibrium Existence and Uniqueness]\label{thm: equilibria always exist and are unique}
    Let $f$ satisfy Assumption \ref{ass: function assumption} and let Assumption \ref{ass: system properties} hold. Then for each constant $w \in \R^n$, the closed loop \crefrange{eq:system description}{eq:controller u} has a unique equilibrium $(x^0$, $z^0)$, inducing input $u^0$ from \eqref{eq:controller u}, which satisfies \crefrange{eq:system description}{eq:controller u} with $\dot{x}=\dot{z}=0$.
\end{theorem}

In addition to the existence of the unique equilibrium $(x^0$, $z^0)$, we can also show that it is globally asymptotically stable under the following assumption on the control parameters.
\begin{assumption} \label{ass: controller tuning}
    Assume that $a_i p_i > r_i$ and $p_i s_i < 1$ for all $i$, where $a_i$ are the diagonal elements of $A$ in \eqref{eq:system description} and $p_i$, $r_i$, and $s_i$ are the controller gains in \crefrange{eq:controller z}{eq:controller u}.
\end{assumption}
\begin{theorem}[Global Asymptotic Stability] \label{thm:stability}
    Let $f$ satisfy Assumption \ref{ass: function assumption} and let $f(u)+h(u)=u$. Let Assumptions \ref{ass: system properties} and \ref{ass: controller tuning} hold. Then there is a globally asymptotically stable equilibrium for the closed loop \crefrange{eq:system description}{eq:controller u}.
\end{theorem}
\begin{remark}
    The tuning rules of Assumption \ref{ass: controller tuning} are fully decentralized. Each agent $i$ can tune their own controller gains to satisfy $r_i < a_i p_i$ and $s_i < 1/p_i$.
\end{remark}
Let us now focus on the case where the function pair $f(\cdot)$ and $h(\cdot)$ are given by the pair $\sat{\cdot}$ and $\dz{\cdot}$ respectively, motivated by classical anti-windup for saturating controllers. Let $\gamma_i$ be positive scalar weights, and consider the problem of minimizing the weighted sum of all state errors $\sum_{i=1}^n \gamma_i |x_i|$. We can define this problem through the optimization problem
\begin{mini!} 
    {x,v}{\sum_{i=1}^n \gamma_i |x_i| = \onenorm{\Gamma x}}
    {\label{eq:optimal problem}}{\label{eq:optimal cost}}
    \addConstraint{-Ax+Bv+w = 0} \label{eq:optimization equality}
    \addConstraint{-\ones \leq v\leq \ones.} \label{eq:optimization constrained u}
\end{mini!}
where $\Gamma = \text{diag}\{\gamma_1, \dots, \gamma_n \}$. The inequalities \eqref{eq:optimization constrained u} are considered componentwise. This problem can be motivated by a district heating example. Let $w$ be the outdoor temperature, $x_i$ be the deviation from the comfort temperature for each agent $i$, and let $B v$ denote the heat provided to the agents, limited by \eqref{eq:optimization constrained u}. Then if $\Gamma = I$, this corresponds to minimizing the total discomfort experienced by all agents. One could consider $\gamma_i$ to be a cost describing the severity of agent $i$ deviating from the comfort temperature, where $\gamma_i$ would be high for e.g. a hospital. Note that this cost does not capture the notion of \textit{fairness} as considered in \cite{agner_coortinating_aw}. For instance, with $\Gamma = I$, $x = \left[ n, 0, \dots, 0 \right]^\top$, and $y = \left[ 1, 1, \dots, 1 \right]^\top$ we achieve the same costs $\onenorm{\Gamma x} = \onenorm{\Gamma y}$. With the problem \eqref{eq:optimal problem} defined, the following holds.

\begin{theorem}[Equilibrium Optimality] \label{thm: optimal equilibrium}
Let Assumption \ref{ass: system properties} hold and let $\Gamma A\inv B$ be a strictly diagonally column-dominant M-matrix. Let $f(u) = \sat{u}$ and $h(u) = \dz{u} = u - \sat{u}$. Let $(x^0$, $z^0)$, be an equilibrium for the closed loop system in \crefrange{eq:system description}{eq:controller u}, associated with input $u^0$. Then $x^* = x^0$ and $v^* = f(u^0)$ solves \eqref{eq:optimal problem}.
\end{theorem}
\begin{remark}\review{
    For an arbitrary choice of $\gamma$, it does not necessarily hold that $\Gamma A\inv B$ is strictly diagonally column-dominant. Thus Theorem \ref{thm: optimal equilibrium} cannot be used as a design-method where we first fix the weights $\gamma$ according to some performance criterion and then calculate a correspondingly optimal controller. However, the set of weights $\gamma_i$ such that the required condition is satisfied will always be non-empty. We can see this because $A$ is a positive definite diagonal matrix by Assumption \ref{ass: system properties} and thus $A\inv B$ is also an M-matrix by Proposition \ref{prop:M matrices} (ii). Proposition \ref{prop:M matrices} (iii) then shows that a positive definite diagonal matrix $\Gamma$ such that $\Gamma A\inv B$ is strictly diagonally column-dominant must exist. Thus Theorem \ref{thm: optimal equilibrium} yields a non-empty set of performance criteria for which the fully decentralized control strategy cannot be outperformed by a more complex control architecture.}
\end{remark}

\section{Numerical Example} \label{sec: example}
This motivating example compares three different control strategies on a simplified, linear model of 10 buildings connected in a district heating grid. The compared strategies are the same as the ones considered in \cite{agner_coortinating_aw}. Each building $i$ has identical thermodynamics on the form
\begin{equation}
    \dot{x}_i =  - \frac{a_i}{C_i} (x_c + x_i - T_\text{ext}(t)) + \frac{1}{C_i}\dot{Q}_i(u),
\end{equation}
where $x_i$ denotes agent $i$'s indoor temperature deviation from the comfort temperature $x_c$, $C_i$ is the heat capacity of each building and $T_\text{ext}$ is the outdoor temperature. $\dot{Q}_i$ is the heat supplied to building $i$. This heat supply is given by
\begin{equation}
    \Dot{Q} = B \sat{u},
\end{equation}
where $B$ represents the network interconnection. The simulation was conducted with $a_i = 0.167\left[ \text{kW/C}^\circ\right]$, $C_i = 2.0 \left[ \text{kWh/C}^\circ\right]$, $p_i = 2.5\left[ \text{1/C}^\circ\right]$, $r_i = 0.2\left[ \text{1/C}^\circ\text{h}\right]$, and $s_i=2.0\left[ \text{C}^\circ\right]$ for all $i$. The parameters $a_i$, $C_i$ are chosen close to the values found in \cite{bacher_identifying_2011} which discusses parameter estimation for a single-family building. \review{The Matrix $B$ is selected as $B_{i,i} = 12$ $\forall i$, $B_{i,j} = -0.15 \mathrm{min}(i,j)$ $\forall i \neq j$ in units $\left[ \text{kW}\right]$.} Matrix $B$ is constructed such that fully opened control valves ($\sat{u} = \ones$) gives $\dot{Q}$ representing a reasonable peak heat demand for small houses. In this scenario, $\dot{Q}_i$ is high for buildings with $i$ small (close to the production facility). We simulate the system using the \texttt{DifferentialEquations} toolbox in Julia \cite{rackauckas2017differentialequations}, for an outdoor temperature scenario given by data from the city of Gävle, Sweden in October 2022 during which the temperature periodically drops to almost -20$^\circ$C. The data is gathered from the Swedish Meteorological and Hydrological Institute (SMHI). \review{This slowly time-varying disturbance also brings insight into how our proposed controller handles a signal $w$ which is not constant.} We compare three different controllers and three different cost functions. The first controller is the fully \textit{decentralized} PI-controller considered in this paper. Secondly the \textit{coordinating} controller consists of the same PI-controllers as the decentralized case, but with the coordinating rank-1 anti-windup signal $\dot{z}_i = x_i + \beta \ones^\top \dz{u}$ considered in \cite{agner_coortinating_aw}. Finally, the \textit{static} controller is given by $u = -B^\top C\inv x$ as considered in \cite{BAUSO_asymptotical_optimality}, where $C$ is the diagonal matrix of all heat capacities $C_i$.

\begin{figure*}[t!]  
    \centering
    \begin{subfigure}{0.45\linewidth}
        \centering
        \begin{tikzpicture}
        \begin{axis}[
            width=\linewidth,
            xlabel={time [h]},
            ylabel={x [$^\circ$C]},
            no markers,
            every axis plot/.append style={blue, solid},
            xmin=0,          
            xmax=336,         
            ymin=-3,
            ymax=0.5,
            grid=both
        ]
        \foreach \i in {1,...,10} {
            \addplot[blue, solid] table [x=t, y=x\i, col sep=comma] {data/x-diagonal.csv};
        }
        \end{axis}
        \begin{axis}[
            width=\linewidth,
            axis y line*=right, 
            axis x line=none,  
            ymin=-20, 
            ymax=15, 
            xmin=0,          
            xmax=336,         
            yticklabel style={color=black}, 
            ylabel style={color=black}, 
        ]
        \addplot[black, no markers, thick, dashed] table [x=t, y=t_out, col sep=comma] {data/t.csv};
        \end{axis}
        \end{tikzpicture}
        \caption{Decentralized controller.}
        \label{fig:decentralized results}
    \end{subfigure}
    \hfill
    \begin{subfigure}{0.45\linewidth}
        \centering
        \begin{tikzpicture}
        \begin{axis}[
            width=\linewidth,
            xlabel={time [h]},
            ylabel={x [$^\circ$C]},
            no markers,
            every axis plot/.append style={blue, solid},
            xmin=0,          
            xmax=336,         
            ymin=-3.0,
            ymax=0.5,
            grid=both
        ]
        \foreach \i in {1,...,10} {
            \addplot[blue, solid] table [x=t, y=x\i, col sep=comma] {data/x-coordinating.csv};
        }
        \end{axis}
        \begin{axis}[
            width=\linewidth,
            axis y line*=right, 
            axis x line=none,  
            ymin=-20, 
            ymax=15, 
            xmin=0,          
            xmax=336,         
            yticklabel style={color=black}, 
            ylabel style={color=black}, 
        ]
        \addplot[black, no markers, thick, dashed] table [x=t, y=t_out, col sep=comma] {data/t.csv};
        \end{axis}
        
        \end{tikzpicture}
        \caption{Coordinating controller.}
        \label{fig:coordinating results}
    \end{subfigure}
    
    \vspace{.5cm}
    
    \begin{subfigure}{0.45\linewidth}
        \centering
        \begin{tikzpicture}
        \begin{axis}[
            width=\linewidth,
            xlabel={time [h]},
            ylabel={x [$^\circ$C]},
            no markers,
            every axis plot/.append style={blue, solid},
            xmin=0,          
            xmax=336,         
            ymin=-3,
            ymax=0.5,
            grid=both
        ]
        \foreach \i in {1,...,10} {
            \addplot[blue, solid] table [x=t, y=x\i, col sep=comma] {data/x-static.csv};
        }
        \end{axis}
        \begin{axis}[
            width=\linewidth,
            axis y line*=right, 
            axis x line=none,  
            ymin=-20, 
            ymax=15, 
            xmin=0,          
            xmax=336,         
            yticklabel style={color=black}, 
            ylabel style={color=black}, 
        ]
        \addplot[black, no markers, thick, dashed] table [x=t, y=t_out, col sep=comma] {data/t.csv};
        \end{axis}
        \end{tikzpicture}
        \caption{Static controller.}
        \label{fig:static results}
    \end{subfigure}
    \caption{Temperature deviations $x$ (blue, left axis) for each strategy and the outdoor temperature $w$ (black, dotted, right axis). Around hour 100, $w$  becomes critically low and the indoor temperatures drop as the controllers saturate.} 
    \label{fig: result time series plots}
\end{figure*}
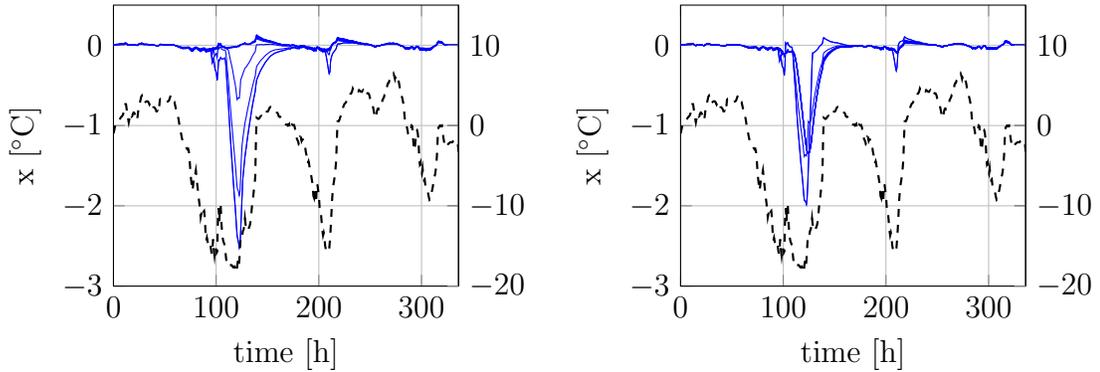
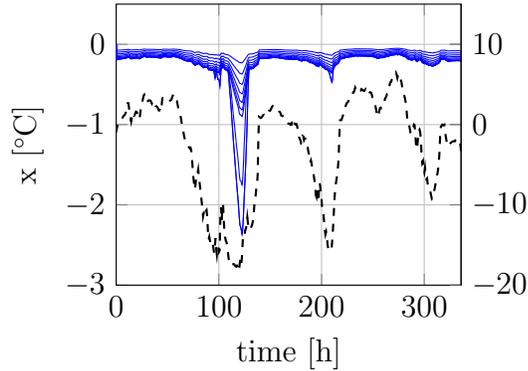

Figure \ref{fig: result time series plots} shows the resulting deviations $x$ during the simulations. At around hour 100, the outdoor temperature is critically low. At this time, the buildings do not receive sufficient heat, regardless of the control strategy. Figure \ref{fig:decentralized results} shows that with the decentralized strategy, the worst deviations become larger than with the coordinating strategy (Figure \ref{fig:coordinating results}). However, not all buildings experience temperature deviations, whereas with the coordinating strategy, all the buildings share the discomfort. Lastly, the static controller has large deviations experienced by many buildings. Even when the outdoor temperature is manageable, the static controller has a constant offset from the comfort temperature, highlighting the usefulness of the integral action. We evaluate the performance through the cost functions
\begin{align}
    J_1 &= \frac{1}{T}\int_0^T \onenorm{x(t)} dt, \label{eq:J1} \\
    J_\infty &= \frac{1}{T}\int_0^T \infnorm{x(t)} dt, \label{eq:Jinf}\\
    J_2 &= \frac{1}{T}\int_0^T x(t)^\top L x(t) + \sat{u(t)}^\top \sat{u(t)} dt. \label{eq:J2}
\end{align}
where $T$ is the simulation time and $L$ is a diagonal matrix where each element is given by $l_i = \frac{q_i}{C_i}$. The cost $J_1$ mimics the optimality notion considered in this paper, $J_\infty$ mimics the optimality notion considered in \cite{agner_coortinating_aw}, and $J_2$ mimics the optimality considered in \cite{BAUSO_asymptotical_optimality}. Table \ref{tab:costs} shows the resulting evaluations. The coordinating controller gives minimal worst-case deviations $J_\infty$, but $J_1$ is minimized in the decentralized strategy. This result, i.e. that the total discomfort is minimized by decentralized control but the worst-case discomfort is minimized by coordination, is found also in \cite{AGNER2022100067} where a nonlinear model of the grid hydraulics and a 2-state model of building dynamics is employed. On the weighted cost $J_2$, all controllers provide similar performance. The static controller slightly outperforms the other two in this scenario, but it is outperformed in every other measure.

\begin{table}[b]
    \centering
    \caption{Costs \crefrange{eq:J1}{eq:J2} evaluated over the simulation.}
    \begin{tabular}{c||c|c|c}
         & Decentralized & Coordinating & Static \\ \hline
         $J_\infty$ & 0.17 & 0.13 & 0.28 \\
         $J_1$ & 0.67 & 0.9 & 1.96 \\
         $J_2$ & 3.52 & 3.52 & 3.49
    \end{tabular}
    \label{tab:costs} 
\end{table}

\section{Proof of Equilibrium Existence and Uniqueness} \label{sec: equilibrium existence}
We will now prove Theorem \ref{thm: equilibria always exist and are unique} through the use of Banach's fixed-point theorem \cite{Agarwal2018}. This proof requires the following two lemmas, the proofs of which are found in the appendix.
\begin{lemma}\label{lem: f shifted has same properties}
    Let $f: \R^n \to \R^n$ and $h: \R^n \to \R^n$ where $h(x) = x - f(x)$ satisfy Assumption \ref{ass: function assumption}. Then $\Tilde{f}: \R^n \to \R^n$ and $\Tilde{h}: \R^n \to \R^n$ given by $\Tilde{f}(x) = f(x + x^0) - f(x^0)$ and $\Tilde{h}(x) = h(x + x^0) - h(x^0)$ for some $x^0 \in \R^n$ also satisfy Assumption \ref{ass: function assumption} and $\Tilde{h}(x) + \Tilde{f}(x) = x$.
\end{lemma}

\begin{lemma}\label{lem: diagonally scaled f}
     Let $f: \R^n \to \R^n$ and $h: \R^n \to \R^n$ where $h(x) = x - f(x)$ satisfy Assumption \ref{ass: function assumption}. Then $\Tilde{f}: \R^n \to \R^n$ and $\Tilde{h}: \R^n \to \R^n$ given by $\Tilde{f}(x) = D f(D\inv x)$ and $\Tilde{h}(x) = D h(D\inv x)$ where $D$ is a diagonal positive definite matrix also satisfy Assumption \ref{ass: function assumption} and $\Tilde{h}(x) + \Tilde{f}(x) = x$.
\end{lemma}

\renewcommand{\proofname}{\textbf{Proof of Theorem \ref{thm: equilibria always exist and are unique}}}
\begin{proof}
    Denote by $S$ a diagonal matrix gathering the positive anti-windup gains $s_i$, $i=1,\dots n$. We can rearrange \crefrange{eq:system description}{eq:controller u} by imposing $\dot{x} = \dot{z} = 0$, which yields
    \begin{equation}
        0 = h(u^0) + S\inv A\inv B f(u^0) + S\inv A\inv w.
        \label{eq: stationary condition}
    \end{equation}
    If there is a unique $u^0$ solving \eqref{eq: stationary condition} then $x^0=A\inv\left(Bf(u^0)+w\right)$ and $z^0 = R\inv(-Px^0 - u^0)$ are uniquely determined by \eqref{eq:system description} and \eqref{eq:controller u} respectively, where $R=\text{diag}\{r_1,\dots,r_n\}$ is invertible by Assumption \ref{ass: controller tuning}. Hence we need only show that there is a unique $u^0$ solving \eqref{eq: stationary condition} for the proof to be complete. Let $D$ be a diagonal positive definite matrix such that $D  S\inv A\inv B D\inv$ is strictly diagonally column-dominant. \review{Note that such a $D$ always exists by Proposition \ref{prop:M matrices} (iv) because $A$ and $S$ are diagonal positive definite and $B$ is an M-matrix.} Left-multiply $\eqref{eq: stationary condition}$ by $D$ and insert multiplication by $I = D\inv D$ before $f(u^0)$ to obtain
    \begin{equation}
        0 = Dh(u^0) + D S\inv A\inv B D\inv D f(u^0) + D S\inv A\inv w.
        \label{eq:banach mid-proof change of vars}
    \end{equation}
    Introduce the change of variables $\hat{B} = D S\inv A\inv B D\inv$, $\zeta = D u^0$, and \mbox{$\hat{w} = D S\inv A\inv w$}. Then \eqref{eq:banach mid-proof change of vars} yields
    \begin{equation}
        0 = Dh(D\inv \zeta) + \hat{B} D f(D\inv \zeta) + \hat{w}.
        \label{eq: equilibrium proof middle-equation}
    \end{equation}
    Here we can use Lemma \ref{lem: diagonally scaled f} to replace $f(\zeta)$, $h(\zeta)$ with $\hat{f}(\zeta) = D f(D\inv \zeta)$, $\hat{h}(\zeta) = Dh(D\inv \zeta)$. Note that $\hat{f}(\cdot)$, $\hat{h}(\cdot)$ satisfy Assumption \ref{ass: function assumption} and $\hat{f}(\zeta) + \hat{h}(\zeta) = \zeta$. Introduce a scalar $k$ satisfying $k > \max( 1, 2 \underset{i}{\max} \hat{B}_{i,i} )$. Divide \eqref{eq: equilibrium proof middle-equation} by $-k$, add $\zeta$ to the left-hand side, and $\zeta = \hat{f}(\zeta) + \hat{h}(\zeta)$ to the right-hand side of \eqref{eq: equilibrium proof middle-equation} to obtain
    \begin{equation}
        \zeta = -\frac{1}{k}\left( (1-k)\hat{h}(\zeta) + (\hat{B} - k I)\hat{f}(\zeta) + \hat{w} \right). \label{eq:banach fixed-point condition}
    \end{equation}
    We define the right-hand side of this expression as $T_w(\zeta)$, defined for a specific $w$. By showing that $T_w$ is a contractive mapping for any $\hat{w}$, we can use Banach's fixed point theorem \cite{Agarwal2018} to show that there is a unique solution $\zeta = T_w(\zeta)$ (and thus a unique $u^0=D\inv \zeta$) for any $\hat{w}$ (and thus any $w = A S D\inv \hat{w}$). Consider any $\alpha \in \R^n$, $\beta \in \R^n$. Then
    \begin{equation}
        T_w(\alpha)-T_w(\beta) =   \frac{-1 + k}{k}\left(\hat{h}(\alpha) - \hat{h}(\beta)\right)  
          + \frac{-\hat{B} + k I}{k}\left(\hat{f}(\alpha) - \hat{f}(\beta)\right) .
    \end{equation}
    Here we use Lemma \ref{lem: f shifted has same properties} to introduce $\Tilde{h}(\alpha-\beta) = \hat{h}(\alpha) - \hat{h}(\beta)$ and $\Tilde{f}(\alpha-\beta) = \hat{f}(\alpha) - \hat{f}(\beta)$. Denote $\Delta = \alpha - \beta$ and $\Delta^+ = T_w(\alpha) - T_w(\beta)$. Then
    \begin{equation}
        |\Delta^+_i| \leq \frac{k - 1}{k}|\Tilde{h}_i(\Delta_i)| + \frac{k - \hat{B}_{i,i}}{k} |\Tilde{f}_i(\Delta_i)| + \sum_{j \neq i}\frac{|\hat{B}_{i,j}|}{k}|\Tilde{f}_j(\Delta_j)|. 
    \end{equation}
    Therefore
    \begin{align}
        \onenorm{\Delta^+} &= \sum_{i=1}^n |\Delta_i^+| \leq \sum_{i=1}^n \bigg( \frac{k - 1}{k}|\Tilde{h}_i(\Delta_i)| \nonumber \\
        & + \frac{k - \hat{B}_{i,i}}{k} |\Tilde{f}_i(\Delta_i)| + \sum_{j \neq i}\frac{|\hat{B}_{j,i}|}{k}|\Tilde{f}_i(\Delta_i)| \bigg).
    \end{align}
    Due to the diagonal column-dominance of $\hat{B}$ and the definition of $k$, it holds that $k > \hat{B}_{i,i} > \sum_{j \neq i}|\hat{B}_{j,i}|$. Thus, selecting $\lambda = \frac{k - 1}{k} < 1$, $\mu_i = \frac{k- \left(\hat{B}_{i,i} - \sum_{j \neq i}|\hat{B}_{j,i}|\right)}{k} < 1$, $\gamma_i = \max(\lambda, \mu_i) < 1$, and $\Bar{\gamma} = \max_i \gamma_i < 1$, we obtain
    \begin{align}
        \onenorm{\Delta^+} &\leq \sum_{i=1}^n \lambda |\Tilde{h}_i(\Delta_i)| + \mu_i |\Tilde{f}_i(\Delta_i)| \nonumber \\
        &\leq \sum_{i=1}^n \gamma_i \left(|\Tilde{h}_i(\Delta_i)| + |\Tilde{f}_i(\Delta_i)| \right) \nonumber \\
        &\leq \sum_{i=1}^n \Bar{\gamma}|\Delta_i| = \gamma \onenorm{\Delta}.
    \end{align}
    Note that $|\Tilde{h}_i(\Delta_i)| + |\Tilde{f}_i(\Delta_i)| = | \Delta_i|$ since $\Tilde{f}_i(\Delta_i)$ and $\Tilde{h}_i(\Delta_i)$ always have the same sign by Assumption \ref{ass: function assumption}, and sum to $\Delta_i$. This proves that $T_w$ is a contraction mapping with respect to the metric $\onenorm{\cdot}$. Thus, by Banach's fixed point theorem, for each $w$ and the ensuing $\hat{w} = D S\inv A\inv w$ there is a unique $\zeta$ such that \eqref{eq:banach fixed-point condition} holds, and thus a $u^0 = D \inv \zeta$ such that \eqref{eq: stationary condition} holds, which completes the proof. 
\end{proof}

\section{Proof of Global Asymptotic Stability}\label{sec: stability}
Given the existence of an equilibrium $(x^0, z^0)$ and the associated input $u^0$, consider the change of variables $\Tilde{z} = - R (z - z^0)$, $\Tilde{u} = u - u^0$, $\Tilde{f}(\Tilde{u}) = f(u^0 + \Tilde{u}) - f(u^0)$, and $\Tilde{h}(\Tilde{u}) = h(u^0 + \Tilde{u}) - h(u^0)$. Due to Lemma \ref{lem: f shifted has same properties}, $\Tilde{f}(\cdot)$, $\Tilde{h}(\cdot)$ satisfy Assumption \ref{ass: function assumption}, and $\Tilde{f}(\Tilde{u})+\Tilde{h}(\Tilde{u}) = \Tilde{u}$. This allows rewriting the \crefrange{eq:system description}{eq:controller u} as
\begin{equation}
\begin{bmatrix}
    \dot{\Tilde{z}} \\
    \dot{\Tilde{u}}
\end{bmatrix} = \begin{bmatrix}
    -R P\inv & R P\inv \\
    A - R P\inv & -A + R P\inv
\end{bmatrix} \begin{bmatrix}
    \Tilde{z} \\
    \Tilde{u}
\end{bmatrix}
- \begin{bmatrix}
    0 \\
    PB
\end{bmatrix} \Tilde{f}(\Tilde{u})
- \begin{bmatrix}
    R S \\
    R S
\end{bmatrix} \Tilde{h}(\Tilde{u})
\label{eq:transformed system}
\end{equation}
where $P$, $R$, and $S$ are diagonal matrices gathering the controller parameters $p_i$, $r_i$, and $s_i$. Stabilizing this system to $\Tilde{z} = \Tilde{u} = 0$ is equivalent to stabilizing the original system system to the equilibrium $x = x^0$, $z = z^0$, and $u = u^0$. We will therefore now prove Theorem \ref{thm:stability} with a Lyapunov-based argument considering system \eqref{eq:transformed system}.

\renewcommand{\proofname}{\textbf{Proof of Theorem \ref{thm:stability}}}
\begin{proof}
Consider the Lyapunov function candidate
\begin{align}
    V(\Tilde{z}, \Tilde{u}) &= \sum_{i=1}^n \int_{0}^{\Tilde{z}_i} q_i (a_i \frac{p_i}{r_i} - 1)\left( \Tilde{f}_i(\zeta) + \epsilon \zeta) \right) d\zeta \nonumber \\
    & + \sum_{i=1}^n \int_{0}^{\Tilde{u}_i} q_i \left( \Tilde{f}_i(\zeta) + \epsilon \zeta \right) d \zeta
\end{align}
where scalars $q_i > 0$ and $\epsilon > 0$ are parameters to be fixed later. For any such choice of parameters, $V$ is positive definite and radially unbounded because $\Tilde{f}_i(\zeta) + \epsilon \zeta$ is increasing in $\zeta$ and zero at zero. Also $a_i \frac{p_i}{r_i} - 1 > 0$ due to Assumption \ref{ass: controller tuning}. The time derivative of $V$ along the trajectories of system \eqref{eq:transformed system} is given by
\begin{subequations}
    \begin{align}
        \Dot{V}(\Tilde{z}, \Tilde{u}) = &-\left(\Tilde{f}(\Tilde{z})+\epsilon \Tilde{z}-\Tilde{f}(\Tilde{u})-\epsilon \Tilde{u}\right)^\top \Tilde{D}(\Tilde{z}-\Tilde{u}) \label{eq: Vdot 1} \\
        & -\left(\Tilde{f}(\Tilde{z})+\epsilon \Tilde{z}\right)^\top \Tilde{D}PS \Tilde{h}(\Tilde{u}) \label{eq: Vdot 2} \\
        & -\left(\Tilde{f}(\Tilde{u})+\epsilon \Tilde{u}\right)^\top QRS \Tilde{h}(\Tilde{u}) \label{eq: Vdot 3} \\
        & -\left(\Tilde{f}(\Tilde{u})+\epsilon \Tilde{u}\right)^\top QPB \Tilde{f}(\Tilde{u}) \label{eq: Vdot 4} 
    \end{align}
\end{subequations}
where $\Tilde{D}$ is a diagonal matrix gathering the positive elements $q_i\left(a_i - r_i/p_i\right)$ and $Q$ is a diagonal matrix gathering the positive elements $q_i$. To simplify this expression, we split it into
\begin{equation}
    \Dot{V}(\Tilde{z}, \Tilde{u}) = \Dot{V}_1(\Tilde{z}, \Tilde{u}) + \Dot{V}_2(\Tilde{z}, \Tilde{u})
\end{equation}
where $\Dot{V}_1(\Tilde{z}, \Tilde{u})$ corresponds to the terms \crefrange{eq: Vdot 1}{eq: Vdot 2} and $\Dot{V}_2(\Tilde{z}, \Tilde{u})$ corresponds to the terms \crefrange{eq: Vdot 3}{eq: Vdot 4}. Since $\Tilde{D}$ and $\Tilde{D}PS$ are diagonal, $\dot{V}_1$ can be analyzed for each $i$ individually. $\Tilde{f}_i(\zeta_i) + \epsilon \zeta_i$ is increasing in $\zeta_i$, therefore $\text{sign}\left(\Tilde{f}_i(\Tilde{z}_i)+\epsilon \Tilde{z}_i-\Tilde{f}_i(\Tilde{u}_i)-\epsilon \Tilde{u}_i\right) = \text{sign}\left(\Tilde{z}_i - \Tilde{u}_i\right)$ and thus \eqref{eq: Vdot 1} is negative semi-definite. If $\Tilde{z}_i$ and $\Tilde{u}_i$ have the same sign, \eqref{eq: Vdot 2} contributes negatively to $\dot{V}_1$. If they have opposite signs the contribution is positive, but then \eqref{eq: Vdot 1} only comprises negative terms as \mbox{$\left(\Tilde{f}_i(\Tilde{z}_i)+\epsilon \Tilde{z}_i-\Tilde{f}_i(\Tilde{u}_i)-\epsilon \Tilde{u}_i\right)\Tilde{D}_{i,i}\left( \Tilde{z}_i - \Tilde{u}_i\right)$} \mbox{$ = \left(|\Tilde{f}_i(\Tilde{z}_i)+\epsilon \Tilde{z}_i|+|\Tilde{f}_i(\Tilde{u}_i)-\epsilon \Tilde{u}_i|\right)\Tilde{D}_{i,i}\left( |\Tilde{z}_i| + | \Tilde{u}_i|\right)$}. Indeed, since $p_i s_i < 1$ from Assumption \ref{ass: controller tuning} and $|\Tilde{h}_i(\Tilde{u}_i)| \leq |\Tilde{u}_i|$ from Assumption \ref{ass: function assumption}, then \eqref{eq: Vdot 1} as developed above dominates \eqref{eq: Vdot 2} which is upper bounded by \mbox{$|\Tilde{f}_i(\Tilde{z}_i) + \epsilon \Tilde{z}_i | \Tilde{D}_{i,i} |\Tilde{h}_i(\Tilde{u}_i)|$}. Thus $\dot{V}_1$ is negative semi-definite. We now turn our attention to $\dot{V}_2$. Note that $\Tilde{u}$, $\Tilde{f}(\Tilde{u})$, and $\Tilde{h}(\Tilde{u})$ elementwise have the same sign and $QRS$ is diagonal, positive definite. Thus 
\begin{align}
\begin{aligned}
&\left(\Tilde{f}(\Tilde{u})+\epsilon \Tilde{u}\right)^\top QRS \Tilde{h}(\Tilde{u}) = \left(\Tilde{f}(\Tilde{u})+\epsilon \Tilde{f}(\Tilde{u}) + \epsilon\Tilde{h}(\Tilde{u})\right)^\top QRS \Tilde{h}(\Tilde{u}) \\
&= (1 + \epsilon)\Tilde{f}(\Tilde{u})^\top QRS \Tilde{h} + \epsilon \Tilde{h}(\Tilde{u})^\top QRS \Tilde{h}(\Tilde{u}) \geq \epsilon \beta \twonorm{\Tilde{h}(\Tilde{u})}^2
\end{aligned}
\label{eq:luca lol}
\end{align}
where $\beta$ is the minimum diagonal element of $QRS$. Note also that 
\begin{align}
\left(\Tilde{f}(\Tilde{u})+\epsilon \Tilde{u}\right)^\top QPB \Tilde{f}(\Tilde{u}) &= (1+\epsilon)\Tilde{f}(\Tilde{u})^\top QPB \Tilde{f}(\Tilde{u}) \nonumber\\
&\quad + \epsilon \Tilde{h}(\Tilde{u})^\top QPB \Tilde{f}(\Tilde{u}). \label{eq:Vdot 4 partitioned}
\end{align}
Fix now the weights $q_i$ in such a way that $QPB + B^\top P Q$ is positive definite. \review{This is possible by Proposition \ref{prop:M matrices} (iv) because $B$ is an M-matrix according to Assumption \ref{ass: system properties}.} Therefore $\exists \alpha > 0$ such that $QPB + B^\top P Q \succ 2 \alpha I$. Thus the first term of \eqref{eq:Vdot 4 partitioned} satisfies
\begin{equation}
    (1+\epsilon)\Tilde{f}(\Tilde{u})^\top QPB \Tilde{f}(\Tilde{u}) \geq (1+\epsilon)\alpha \twonorm{\Tilde{f}(\Tilde{u})}^2.
    \label{eq:Vdot 4 bound on fBf}
\end{equation}
We also note that the second term in \eqref{eq:Vdot 4 partitioned} satisfies
\begin{equation}
    \epsilon\Tilde{h}(\Tilde{u})^\top QPB \Tilde{f}(\Tilde{u}) \geq -\epsilon\gamma \twonorm{\Tilde{f}(\Tilde{u})}\twonorm{\Tilde{h}(\Tilde{u})}
    \label{eq:Vdot 4 bound on hBf}
\end{equation}
where $\gamma = \twotwonorm{QPB}$. Thus, combining the bounds in \eqref{eq:luca lol}, \eqref{eq:Vdot 4 bound on fBf} and \eqref{eq:Vdot 4 bound on hBf} within \crefrange{eq: Vdot 3}{eq: Vdot 4}, we obtain
\begin{equation}
\begin{aligned}
    \Dot{V}_2(\Tilde{z}, \Tilde{u}) &\leq -(1+\epsilon)\alpha\twonorm{\Tilde{f}(\Tilde{u})}^2 - \epsilon \beta \twonorm{\Tilde{h}(\Tilde{u})}^2 + \epsilon \gamma \twonorm{\Tilde{f}(\Tilde{u})}\twonorm{\Tilde{h}(\Tilde{u})} \\
    &=
    \begin{pmatrix}
        \twonorm{\Tilde{f}(\Tilde{u})} \\ \twonorm{\Tilde{h}(\Tilde{u})} 
    \end{pmatrix}^\top
    \begin{pmatrix}
        -(1+\epsilon)\alpha & \frac{1}{2}\epsilon \gamma \\
        \frac{1}{2}\epsilon \gamma & - \epsilon \beta
    \end{pmatrix}
    \begin{pmatrix}
        \twonorm{\Tilde{f}(\Tilde{u})} \\ \twonorm{\Tilde{h}(\Tilde{u})} 
    \end{pmatrix}.
\end{aligned}
\label{eq:Vdot_2}
\end{equation}
We may now select the Lyapunov function parameter $\epsilon$ sufficiently small such that $\left(\alpha + \epsilon\alpha - \frac{\epsilon \gamma^2}{4 \beta}\right) > 0$. This makes the quadratic form \eqref{eq:Vdot_2} negative definite. Thus $\Dot{V}_2(\Tilde{z}, \Tilde{u})=0$ if and only if $\Tilde{f}(\Tilde{u}) = \Tilde{h}(\Tilde{u}) = 0$, i.e. if and only if $\Tilde{u} = 0$. In this case, $\Dot{V}_1(\Tilde{z}, \Tilde{u})$ is clearly negative definite in $\Tilde{z}$. Thus $\Dot{V}(\Tilde{z}, \Tilde{u})$ is negative definite, which implies that the origin is globally asymptotically stable for system \eqref{eq:transformed system}. Equivalently, the equilibrium $(x^0,z^0)$, with input $u^0$, is therefore globally asymptotically stable for the original system \crefrange{eq:system description}{eq:controller u}.
\end{proof}

\section{Proof of Equilibrium Optimality}\label{sec:optimality proof}
Here we prove Theorem \ref{thm: optimal equilibrium}.
\renewcommand{\proofname}{\textbf{Proof of Theorem \ref{thm: optimal equilibrium}}}
\begin{proof}
    Firstly, it is clear that $v^* = \sat{u^0}$ and $x^*_i = x^0_i = -s_i\dz{u^0_i}$ for all $i$ satisfies \eqref{eq:optimization equality} due to $x^0$, $z^0$ being an equilibrium, and satisfies \eqref{eq:optimization constrained u} because $\sat{\cdot}$ is bounded in the range $\left[-1,1\right]$. Consider, for establishing a contradiction, that there exists $\mu \neq 0$ such that $v^\dagger = v^* + \mu$ and $x^\dagger = A\inv Bv^\dagger + A\inv w = x^*+A\inv B \mu$ is the optimal solution to \eqref{eq:optimal problem} with a smaller cost \eqref{eq:optimal cost} than the one obtained by $x^*$, $v^*$. Then $\mu$ solves the optimization problem
    \begin{mini!} 
        {\mu}{\sum_{i=1}^n |\gamma_i x^*_i + \Tilde{B}_i \mu|}
        {\label{eq:second optimal problem}}{\label{eq:second optimal cost}}
        \addConstraint{-\ones \leq v^* + \mu \leq \ones.} \label{eq:second optimization constrained u}
    \end{mini!}
    where $\Tilde{B}_i$ is row $i$ of the matrix $\Tilde{B} = \Gamma A \inv B$. The equilibrium of \eqref{eq:controller z} implies $x^*_i = -s_i\dz{u^0_i}$. Therefore we can leverage \eqref{eq:second optimization constrained u} to see that $x^*_i > 0 \implies u^0_i < -1 \implies v_i = -1 \implies \mu_i \geq 0$ and conversely $x^*_i < 0 \implies u^0_i > 1 \implies v_i = 1 \implies \mu_i \leq 0$. Combining this with $\Gamma$ and $A$ both being diagonal, positive definite and the fact that $B$ is an M-matrix which implies that $\Tilde{B}_{i,i} > 0$, we obtain $|\gamma_i x_i + \Tilde{B}_{i,i} \mu_i| = |\gamma_i x_i| + |\Tilde{B}_{i,i} \mu_i|$ for all $i$. Thus \eqref{eq:second optimal cost} can be expanded as follows
    \begin{equation}
        \begin{aligned}
            \sum_{i=1}^n |\gamma_i x_i + \Tilde{B}_i \mu| &\geq \sum_{i \neq j} \left(|\gamma_i x_i + \Tilde{B}_{i,i} \mu_i| - |\sum_{i\neq j} \Tilde{B}_ {i,j} \mu_j|\right) \\
             &\geq \sum_{i=1}^n \left(|\gamma_i x_i| + |\Tilde{B}_{i,i}| |\mu_i|\right) - \sum_{i=1}^n\sum_{j \neq i} |\Tilde{B}_{i,j}| |\mu_j| \\
             & = \sum_{i=1}^n |\gamma_i x_i| + \sum_{k=1}^n\left(|\Tilde{B}_{k,k}| - \sum_{j\neq k} |\Tilde{B}_{j,k}|\right)|\mu_k|.
        \end{aligned}
    \end{equation}
    Since $\Tilde{B}$ is diagonally column-dominant, then $|\Tilde{B}_{k,k}| - \sum_{j\neq k} |\Tilde{B}_{j,k}|$ is positive for all $k$. Thus this expression is minimized by $\mu = 0$, which completes the proof.
\end{proof}

\section{Conclusions} \label{sec: conclusions}
In this paper we considered fully decentralized PI-control for a class of interconnected systems subject to incrementally sector-bounded nonlinearities. We showed that for systems where the input matrix is an M-matrix, fully decentralized PI-controllers globally asymptotically stabilize a specific equilibrium. Furthermore, this equilibrium is optimal in that it minimizes costs of the form $\sum_{i=1}^n \gamma_i| x_i|$.
The proposed control strategy was employed in a numerical example of a simplified district heating system model. The example showed that, with our decentralized strategy, the total discomfort in the system is minimized, at the cost of higher worst-case discomforts when compared with a alternative coordinated control strategies. We have thus demonstrated that a fully decentralized and easily tuned control law constitutes a relevant design for a large class of systems.

Open questions include analysis of the transient response, and finding controller tuning rules accordingly. \review{This could encompass the case when $w$ is not constant but slowly time-varying, such as in the simulation study in Section \ref{sec: example}}. Furthermore, to better capture the district heating application, a richer class of systems should be considered: Multi-state models for each building, as well as more complex, nonlinear models of the interconnection $B$ can be considered.

\section*{Declaration of Competing Interest}
The authors declare that they have no known competing financial
interests or personal relationships that could have appeared to influence the work reported in this paper.

\section*{Acknowledgements}
Felix Agner, Jonas Hansson and Anders Rantzer are members of the ELLIIT Strategic Research Area at Lund University.

This work is funded by the European Research Council (ERC) under the European Union's Horizon 2020 research and innovation program under grant agreement No 834142 (ScalableControl).

This work was funded by Wallenberg AI, Autonomous Systems and Software Program (WASP) funded by the Knut and Alice Wallenberg Foundation.

Work supported in part by the MUR via grant DOCEAT, CUP E63C22000410001, number 2020RTWES4, and by the ANR via grant OLYMPIA, number ANR-23-CE48-0006.
 \bibliographystyle{elsarticle-num} 
 \bibliography{bibliography}





\appendix
\section*{Appendix}  
\counterwithin{equation}{section}  
\renewcommand{\theequation}{A.\arabic{equation}}  

We prove here suitable properties of the function class characterized by Assumption \ref{ass: function assumption}, as stated in Lemmas 1, 2 and 3. To simplify the exposition, we drop the subscript $i$.
\renewcommand{\proofname}{\textbf{Proof of Lemma \ref{lem: u - f(u) has same properies}}}
\begin{proof}

    Clearly, $h(0) = 0 - f(0) = 0$. Additionally,
    \begin{equation}
        \frac{h(y)-h(x)}{y-x}= \frac{y-f(y) - x + f(x)}{y-x} = 1 - \frac{f(y)-f(x)}{y-x} \in \left[0, 1\right] 
    \end{equation}
    which shows that \mbox{$0 \leq (h(y)-h(x))/(y-x) \leq 1$} if $x \neq y$, concluding the proof. 
\end{proof}
\renewcommand{\proofname}{\textbf{Proof of Lemma \ref{lem: f shifted has same properties}}}
\begin{proof}

    Clearly, $\Tilde{f}(0) = f(x^0)-f(x^0)=0$. In addition,
    \begin{equation}
        \frac{\Tilde{f}(y)-\Tilde{f}(x)}{y-x}= \frac{f(y+x^0)-f(x+x^0)}{(y+x^0)-(x+x^0)} \in \left[0, 1\right]
    \end{equation}
    which shows that $0 \leq (\Tilde{f}(y)-\Tilde{f}(x))/(y-x) \leq 1$ if $x \neq y$. Finally $\Tilde{f}(x) + \Tilde{h}(x) = f(x + x^0) - f(x^0) + h(x + x^0) - h(x^0) = x + x^0 - x^0 = x$, concluding the proof. 
\end{proof}
\renewcommand{\proofname}{\textbf{Proof of Lemma \ref{lem: diagonally scaled f}}}
\begin{proof}

    $\Tilde{f}(0) = D\inv f(0) = 0$. Additionally,
    \begin{equation} 
        \frac{\Tilde{f}(y)-\Tilde{f}(x)}{y-x} = \frac{df(y/d)-df(x/d)}{y-x}= \frac{f(y/d)-f(x/d)}{y/d-x/d} \in \left[0, 1\right]. 
    \end{equation}
    Thus \mbox{$0 \leq (\Tilde{f}(y)-\Tilde{f}(x))/(y-x) \leq 1$} if \mbox{$x \neq y$}. Finally $\Tilde{f}(x) + \Tilde{h}(x) = \\D f( D\inv x) + D h(D\inv x)= D \left( f( D\inv x) + h(D\inv x)\right) = D D\inv x = x$, concluding the proof. 
\end{proof}

\end{document}